\documentclass[aps, sort&compress, square, floatfix]{revtex4}
\usepackage[dvips]{graphicx}

\begin{document}

\title{Generalised Modal Analysis with the Pad\'e-Laplace transform}

\author{C. Tannous}

\affiliation{Laboratoire de Magn\'etisme de Bretagne,  CNRS/UMR 6135
6 avenue Le Gorgeu BP:  809,  29285 Brest CEDEX,  France}

\begin{abstract}
We present the  G-MAPLE (Generalised Modal Analysis
from the Poles of the Laplace Expansion) software that allows decomposing 
data depending on a single parameter (such as time series data) into a set of exponential 
functions having complex amplitudes and arguments.  The novelty is that 
G-MAPLE determines the unknown number of exponentials in the data along with 
the corresponding complex amplitudes and arguments. 
\end{abstract}

\maketitle

\section{Introduction}
Decomposing a series of data points depending on
a single parameter (that might be time,  temperature,  pressure...) is an
important problem in Science and Technology. \\

The data might be conductivity,  chemical concentration,  
stock market volume,  a communication signal etc...  
It is required that the data is suspected to contain an unknown number 
of exponential components possessing complex amplitudes and having complex 
arguments linear functions of the single parameter at hand \cite{tufts,cadzow,rao}.\\

The important point is that we do not have to specify the number of 
exponential components: the Pad\'e-Laplace algorithm finds it. This means, the algorithm 
performs Fourier analysis in the case of periodic or multiperiodic input 
data, decomposes a purely decaying function into a set of exponentials 
having real arguments in contrast to Fourier analysis where the arguments 
are pure imaginary, and finally analyses a function that decays while 
simultaneously oscillating as consisting of a set of exponentials having 
complex arguments. In all three cases the amplitudes are determined in 
complex form. It is a well-known fact that Fourier/spectral analysis breaks 
down when dealing with exponentially decaying functions  and numerical 
analysis or standard fitting techniques fail when dealing with functions of 
that sort, because of the ill-posed nature of the problem of exponential 
fitting \cite{acton}. G-MAPLE circumvents ill-posedness and succesfully gives, not only 
the number of exponential components in the function, but also the 
corresponding amplitudes and arguments. \\

The applications of Generalised Modal analysis are numerous. Some of
the fields of applications are: 

\begin{enumerate}

\item Numerical fitting software:  \\
Exponential and trigonometric fitting of functions as used in the general
Pade rational approximation approach encountered in Quantum Mechanics \cite{leung},
Particle physics, statistical physics \cite{wittle}....

\item Time series software:   \\
Determination of time constants in arbitrary profiles of use in general 
diagnosis, seasonal trends, forecasting etc...\cite{scharf}.

\item Communications engineering:  \\
Separation of a signal from its echoes in telephone lines, multipath links 
(line of sight microwave propagation or mobile outdoor/ indoor); traffic 
analysis requiring the determination of different parameters yielding 
traffic volume such as mean arrival times at a given channel \cite{sarkar}.

\item Chemical engineering, oil, gas, pharmaceutical...:   \\
Determination of the order of a given chemical reaction, detection of 
intermediate reactions, help in the finding of intermediate reactants or 
processes etc... here, one expects exponential functions of temperature as 
well, due to the various thermal activation or propagation processes \cite{rasol}.

\item Control systems:  \\ 
Detection of the optimal or hidden number of parameters governing the 
behaviour of a complex system or plant... 

\item Mechanical engineering:   \\
Determination of the number of degrees of freedom (linear or angular 
displacements or velocities) in the the response of mechanical system 
such as a robot, mechanical arm, calculation of friction constants controlling 
the motion etc...

\item Population models in economics, biology, ecology...:  \\
Determination of number and type of species (biological or economic 
entities) and understanding the underlying dynamics giving rise to the 
overall observed population or stock market volume at any given time...
For instance, it is used in fluorescence decay studies in Biochemistry
\cite{foucault, scalley}.

\item Speech and acoustic engineering:   \\
Determination of basic components of a phoneme in the time domain, in 
contrast to the windowed spectral analysis method that is usually performed. 
echo and reverberation investigation in concert halls...

\item Geophysical prospecting and seismic data processing:  \\
Separation of the various echoes produced by a shock wave, seism and 
identification of the various types and spatial extent of different layers 
of materials, strata... \cite{mehta}.

\item Micro-electronics industry:   \\
Determination of types of impurities or dopants when one monitors the total 
current versus temperature or the time dependent capacitance in DLTS (Deep 
Level Transient Spectroscopy) studies...
\end{enumerate}

More recently, it was applied to gene sequencing where exponential functions
are used in the calculation of partition functions of DNA loop sequences
\cite{gorban, yeram00}.

\section{Software presentation}

G-MAPLE is a software that decomposes the given data in a set of exponential 
functions having complex amplitudes and arguments. The novelty is that 
G-MAPLE determines the unknown number of exponentials in the data along with 
their complex amplitudes and arguments \cite{yeram87}. \\
G-MAPLE does a Taylor expansion of the Laplace transform of the data around 
a selected point in the complex plane,  fits the expansion to a rational 
approximation,  finds its poles and infers from them the arguments of the 
exponential components and the values of their corresponding amplitudes.  The 
number of exponential components is found by increasing steadily the degree 
of the rational approximation.  Presumably a set of poles will appear beyond 
a certain degree and keep appearing later on,  regardless of the degree.  The 
number of stable poles is the number of exponential components sought.  These 
stable poles and their associated amplitudes give the exponential functions 
the sum of which is the best fit to the initial data.   \\ 

Hence the plan of the G-MAPLE strategy \cite{claverie} is: 
\begin{itemize}
\item Preprocessing of the data:  Filtering/Estimating the noise contained in 
the data or/and altering the sampling rate of the data. 
\item Determination of the optimal point $p_0$ for the Taylor expansion of the 
Laplace transform of the data. 
\item Pad\'e rational approximation to the Taylor expansion. 
\item Inspection of pole stability. 
\item Consistency analysis. 
\end{itemize}

\subsection{Preprocessing of the data:}
This is performed with a spline program.  Once the data is entered an
an estimate of the noise and a smoothing parameter are provided. 
Altering and trying  several sampling rates is done because it is important for 
the accuracy required in the calculation of the Taylor coefficients.
 
\subsection{Determination of the optimal expansion point $p_0$: }
\begin{enumerate}
\item Initial study of the approximate locations of the poles from the 
Laplace transform of the data. 
\item Performing the Taylor expansion for a given degree around several 
expansion points and examining the behaviour of the coefficients with the 
degree. 
\end{enumerate}

The estimation of the Taylor coefficients of the Laplace transform   \cite{claverie} 
of the data is done for several trial p value.  An acceptable value for the 
expansion point $p_0$  should yield a set of coefficients whose magnitude decrease smoothly with 
their algebraic normalised value staying at about the same level.  This is 
done by a simple inspection of the plots of the above sets of coefficients 
versus the various expansion points.  Once the point $p_0$ is found one can 
start the rational approximation analysis.  By the same token the damping 
coefficient is calculated as well.  This is required for rapidly varying 
data.  It is possible sometimes to alter slightly the damping coefficient in 
order to get a better expansion point. 
\subsection{Rational approximation procedure: }
Once the Taylor coefficients of the Laplace transform have been determined 
about the optimal expansion point $p_0$,  one has to vary the degree of the 
rational approximation to the Taylor expansion.  One selects a high 
degree (5...12) and calculates the rational 
approximation for all degrees:  beginning from degree 2 to the highest possible degree. 
One repeats for every degree between 2 and the largest one the calculation of 
the Taylor expansion and the rational approximation. All poles and corresponding
 amplitudes for every degree are calculated and stored them in a 
master pole file. The latter is required for the stability analysis. Regarding this 
procedure,  two comments are in order: 
\begin{enumerate}
\item It is is very important to choose 
carefully the expansion point $p_0$ of the Taylor expansion.  
Once it is chosen the Taylor coefficients will be determined and 
from them the poles of the Laplace transform are obtained along with the 
corresponding amplitudes. 
\item When the poles and amplitudes are found one has to assess the implications 
of the signs of the amplitudes and poles.  For example:  If the data shows a 
monotonic falloff,  then the poles should all be real negative.  If the data 
show oscillatory behaviour,  then one must expect imaginary or complex poles. 
The poles and amplitudes stored in pole. file for all degrees calculated,  
from 2 to the highest one (5..12) will be analysed by a stability check
procedure that will establish the history of occurrence of each pole as the 
degree has been varied. 
\end{enumerate}
\subsection{Pole stability analysis: }
After poles and amplitudes are calculated from the Laplace and rational
approximation,  one has to study the 
number of occurrence of each pole with its corresponding amplitude.  Since we 
are looking at the frequency of occurrence of poles in the two-dimensional 
complex plane,  we have to look at small areas where a given pole keeps 
showing up persistently beyond a certain degree.  For a given  
rough estimate of the size of the area in which the occurrence of poles is 
confined (an interactive test is provided for the size of the radius of 
the circular area).  The sensitivity analysis of the 
presumed stable  poles and their occurrence in that small area is performed
in order to evaluate the robustness of the poles found.
\subsection{Consistency analysis: }
As a final analysis of the soundness of the results,  the software provides 
three capabilities: 
\begin{enumerate}
\item The sequence of ratios of the Taylor coefficients should converge towards 
the smallest modulus of the poles found.  This is a test of the accuracy of 
the calculation of the Taylor coefficients. 
\item Once the stable amplitudes and poles are found and have passed all tests 
descibed above,  one can define an analytical function from them that can 
generate data to be compared with the initial input data. 
\item If the analysis fails or one wishes to perform a robustness analysis,  it 
is required that the Taylor coefficients are well calculated since all the 
subsequent analysis is based on them.  The software allows the selection of 
several integration techniques along with several possible interpolation 
routes for the data to analyse. One may compare the values of the 
coefficients under the different assumptions and choose the satisfactory 
ones. 
\end{enumerate}
\section{Detailed description of the processing steps:}
\begin{enumerate}
\item Preprocessing: \\
Performing the smoothing of the data,  estimating the noise contained in 
it and altering the sampling rate at will for accuracy 
considerations during the calculation of the Taylor coefficients. 
\item Finding the optimal expansion point $p_0$: \\
Two procedures are used: 
  \begin{enumerate}
\item The Laplace transform of the data is performed in order to find the 
approximate locations of the poles.  The optimal expansion point $p_0$ should be larger 
than the largest pole found. 
\item Determination of the optimal point: This performs the Taylor expansion
 of the Laplace transform of the data 
and estimates the damping coefficient required for rapidly varying data 
values.  The Taylor coefficients are plotted as a function of their order for 
the given degree and set of points chosen.
 \end{enumerate} 
\item Rational approximation to the Taylor expansion of the Laplace transform of 
the data: \\ 
This step is to perform the rational approximation and find the poles and 
amplitudes for all degrees starting from 2 up to the highest degree chosen.  
\item Stability Analysis: \\
This step performs the stability analysis.  The stable poles and amplitudes are 
found and stored. 
\item Consistency and final checks: 
\begin{enumerate}
\item Generating an analytical function from the stable poles and 
amplitudes found. 
The accuracy of the Taylor coefficients is being granted by the graph 
showing whether their ratios approch the minimal pole modulus or not as 
their order increases. 
\item Generating data points from the constructed analytical function and adding
synthetic noise. 
One might even add Gaussian or Uniform noise in order to make it as similar 
as possible to the initial data. 
\item Accuracy testing: This consists in obtaining the Taylor coefficients
 with various methods depending 
on the data available and the accuracy needed.  It is advised to see how the various ways 
of calculating the coefficients alters the final results (poles and amplitudes). 
\end{enumerate}
\end{enumerate}

\section{Procedure summary:}
The summary reviews every step and indicates how the required input data is processed   
 and how the results obtained at every step described previously are checked: 
\begin{enumerate}
\item Preprocessing:  The user has to provide the data file name,  the number of 
data points and how the file is organized (data only, index and data or x, y data). 
In the filtering step,  an estimate of the noise 
contained in the data as well as a smoothing parameter should be provided.  The spline procedure 
returns an estimate of the noise and if the smoothing parameter is well 
chosen the initial estimate of the noise and the returned noise figure 
should be roughly in agreement. 
Finally it is possible to choose a different sampling rate of the data once it has been 
smoothed and splined.  
This has implications on the accuracy of the Taylor coefficients and a test exists
 in the software to check 
this accuracy. 
\item Optimal expansion point:  Assuming a range of points about 
which the Taylor coefficients are calculated for a chosen degree, 
the software returns the graphs and values of these coefficients versus the 
degree. 
\item Rational approximation:   Providing an upper value of the degree for 
which the analysis is performed, the software returns all poles and 
amplitudes for all degrees between degree 2 and the largest degree 
chosen.  It returns also a consistency graph showing the sequence of ratios 
of Taylor coefficients for increasing order versus the minimal modulus of 
the poles found. 
\item Stability Analysis:  For a given radius (within bounds) of the circular 
neighborhood about which the frequency of occurrence of a given pole is 
examined, this analysis is done.  
The software  returns the history of occurrences of the poles 
and determines the stable ones. 
\item Consistency Analysis:  This is done automatically through the display of 
the graph of the ratios of the Taylor coefficients versus order with respect 
to the smallest modulus of the poles found with the Pad\'e-Laplace rational
approximation.  A new set of data from the 
stable poles and amplitudes found can be generated and compared it to the 
initial input data.  It is advised to examine the sensitivity of the Taylor coefficients 
with respect to the various ways of calculating them,  depending on the 
type of data available. 

\end{enumerate}
\section{Examples:}
\subsection{Synthetic example: Degree 12 Lanczos}
We start first with a synthetic example.  The parameters are:  \\

Amplitudes: \\
3.00000 \\
0.400000 \\
3.40000 \\
6.20000 \\

Corresponding exponential arguments (poles): \\
(-1.0000000000000,  0.) \\
(-3.0000000000000,  0.) \\
(-5.0000000000000,  0.) \\

Damping parameter: -4.16 \\

\begin{enumerate}

\item Determination of the optimal expansion point and the damping coefficient:  \\
the taylor expansion coefficients c(r) are listed according to their absolute 
values whereas r!c(r) are given in algebraic values:  \\

$p$ = 1.000D-01,   \\ 

c(1)= 2.6760968316948  \\ 
c(2)= -0.92274649709482   \\ 
c(3)= 0.46139353620293  \\ 
c(4)= -0.31918555072345  \\ 
c(5)= 0.26204150623143   \\ 
c(6)= -0.22997288909213   \\ 
c(7)= 0.20656813173070  \\ 
c(8)= -0.18701514280649  \\ 
c(9)= 0.16977029842334  \\ 
c(10)= -0.15425931881287  \\ 
c(11)= 0.14021104327295  \\ 
c(12)= -0.12745666366475  \\ 
c(13)= 0.11586714821751  \\ 
c(14)= -0.10533295150486  \\ 

$$
\begin{array}{rl}
*************************  & c(1)   \\ 
********* & c(2)    \\  
***** & c(3)    \\ 
*** & c(4)   \\  
*** &  c(5)    \\ 
***  & c(6)    \\ 
** & c(7)    \\ 
** & c(8)    \\ 
** & c(9)    \\ 
** & c(10)    \\ 
**  & c(11)    \\ 
** &  c(12)    \\ 
** &  c(13)    \\ 
* & c(14)    \\ 
\end{array}
$$

$$
\begin{array}{rl}
***********************  & 1! c(1)   \\ 
*********************** &  2! c(2)   \\ 
*********************** & 3! c(3)   \\ 
*********************** &  4! c(4)   \\ 
*********************** &  5! c(5)   \\ 
*********************** &  6! c(6)   \\ 
*********************** &  7! c(7)   \\ 
*********************** &  8! c(8)   \\ 
*********************** & 9! c(9)   \\ 
***********************  & 10! c(10)   \\ 
*********************** &  11! c(11)   \\ 
********************** & 12! c(12)   \\ 
*************************  & 13! c(13)   \\ 
* & 14! c(14)   \\ 
\end{array}
$$

Optimal expansion point $p_0$=0.4\\

\item Laplace and rational approximation calculations for degrees= 2...12 \\

\item Stability study of the poles versus radius:  \\ 
 \\   radius=1  \\
 number of stable poles= 8  \\
 pole {\#} 1 (-4.0787128997673,  0.) appeared 11 times  \\
 pole {\#} 2 (-1.0821008803446,  0.) appeared 14 times  \\
 pole {\#} 3 (-3.0000000476863,  0.) appeared 10 times  \\
 pole {\#} 4 (3.8482758680100,  0.) appeared 3 times  \\
 pole {\#} 5 (-8.6468139578141,  0.) appeared 2 times  \\
 pole {\#} 6 (1.8636510844379,  0.) appeared 6 times  \\
 pole {\#} 7 (-6.3269596366764D-02,  -0.66701505012643) appeared 26 times  \\
 pole {\#} 8 (-6.3269590975568D-02,  0.66701507038499) appeared 5 times  \\
 the number of occurrences of some pole is larger than the highest degree = 12  \\

  Going to a smaller radius=10$^{-2}$  \\

 Number of stable poles= 10  \\

 pole {\#} 1 (-4.9999999672223,  0.) appeared 10 times  \\
 pole {\#} 2 (-3.0000000476863,  0.) appeared 10 times  \\
 pole {\#} 3 (-0.99999996721772,  0.) appeared 10 times  \\
 pole {\#} 4 (3.8482758680100,  0.) appeared 2 times  \\
 pole {\#} 5 (-8.6468139578141,  0.) appeared 2 times  \\
 pole {\#} 6 (1.8636510844379,  0.) appeared 2 times  \\
 pole {\#} 7 (-6.3269596366764D-02,  -0.66701505012643) appeared 2 times  \\
 pole {\#} 8 (-6.3269590975568D-02,  0.66701507038499) appeared 2 times  \\
 pole {\#} 9 (2.7049633928318,  -1.6749074721005D-09) appeared 2 times  \\
 pole {\#} 10 (0.40001532007036,  2.9758199321717D-03) appeared 2 times
  \\
\item  Consistency check: 
  \\
 Occurrence of the poles versus degree (1 for occurrence and 0 for no-occurrence):  \\

 pole {\#} 1:  0 1 1 1 1 1 1 1 1 1 1  \\
 pole {\#} 2:  0 1 1 1 1 1 1 1 1 1 1  \\
 pole {\#} 3:  0 1 1 1 1 1 1 1 1 1 1  \\
 pole {\#} 4:  0 0 1 0 0 0 0 0 1 0 0  \\
 pole {\#} 5:  0 0 0 1 0 0 0 1 0 0 0  \\
 pole {\#} 6:  0 0 0 1 0 0 0 1 0 0 0  \\
 pole {\#} 7:  0 0 0 0 1 0 1 0 0 0 0  \\
 pole {\#} 8:  0 0 0 0 1 0 1 0 0 0 0  \\
 pole {\#} 9:  0 0 0 0 1 0 1 0 0 0 0  \\
 pole {\#} 10:  0 0 0 0 0 0 1 0 0 0 0 \\
 
 After reducing radius to 10$^{-5}$  \\

 Number of stable poles= 3  \\
 pole {\#} 1 (-4.9999999672223,  0.) appeared 10 times  \\
 pole {\#} 2 (-3.0000000476863,  0.) appeared 10 times  \\
 pole {\#} 3 (-0.99999996721772,  0.) appeared 10 times
  \\
 pole {\#} 1:  0 1 1 1 1 1 1 1 1 1 1  \\
 pole {\#} 2:  0 1 1 1 1 1 1 1 1 1 1  \\
 pole {\#} 3:  0 1 1 1 1 1 1 1 1 1 1
  \\
 Conclusion of the analysis:   \\
 Rational approximation degree= 12  \\
 Expansion point:  0.400000, Damping parameter= -4.16000  \\

\item  Stable poles and their corresponding amplitudes:   \\

 amplitude {\#} 1 (6.1999983164819,  -9.6235532915790D-10)   \\
 pole {\#} 1 (-5.0000004239243,  -1.1584249121758D-11)  \\
 amplitude {\#} 2 (3.4000016676147,  1.2316899810646D-08)   \\
 pole {\#} 2 (-3.0000003747023,  9.9268993397018D-12)  \\
 amplitude {\#} 4 (0.39999985296666,  -1.8860442419695D-07)   \\
 pole {\#} 4 (-0.99999997587661,  2.7211949994764D-10)
  \\
 for comparison, the input amplitudes and arguments are:   \\

 3  \\

 0.400000  \\
 3.40000  \\
 6.20000  \\

 (-1.0000000000000,  0.)  \\
 (-3.0000000000000,  0.)  \\
 (-5.0000000000000,  0.)  \\

\end{enumerate}

\subsection{Decaying only exponentials}

 5    \\

    (200.000,0.)      \\
   (1000.000,0.)      \\
   (500.000,0.)      \\
    (800.000,0.)      \\
    (300.000,0.)      \\

  (  -50.000000000000,  0.)      \\
  (  -5.0000000000000,  0.)      \\
  ( -0.50000000000000,  0.)      \\
  ( -0.25000000000000,  0.)      \\
  ( -0.15000000000000,  0.)      \\

         Damping parameter=     -5.00     \\

     Expansion point $p_0$=1.2, damping parameter=-5.    \\

   amplitude \#  1  (   200.44842041705,   -6.4297285954932D-08)      \\
   pole      \#  1  (  -50.199847855334,    1.7266006542437D-09)      \\

   amplitude \#  2  (  1000.06460142389,   -5.7036189955559D-06)      \\
   pole      \#  2  (  -5.0002454942104,   -1.3064326900177D-08)      \\

   amplitude \#  3  (   499.40738201272,    1.3497166341200D-05)      \\
   pole      \#  3  ( -0.50015131820020,   -1.6254860197388D-11)      \\

   amplitude \#  4  (   798.94082988133,    2.3349965623979D-05)      \\
   pole      \#  4  ( -0.25024465114282,  0.)      \\

   amplitude \#  5  (   301.65857589825,    9.1799870303999D-06)      \\
   pole      \#  5  ( -0.15016920303216,  0.)      \\

   amplitude \#  6  (  -6.8688938206640D-08,   -3.5775972803453D-08)      \\
   pole      \#  6  (  0.51069582321812,   0.17487380533828)      \\

   amplitude \#  7  (  -6.8686582230124D-08,    3.5777421059967D-08)      \\
   pole      \#  7  (  0.51069582321812,  -0.17487380533828)      \\

   amplitude \#  8  (   3.9828988025503D-05,   -1.2554952329213D-05)      \\
   pole      \#  8  (   2.0290173507300,    1.2045224512252)      \\

   amplitude \#  9  (  -2.3797048952317D-05,   -2.7704251925653D-05)      \\
   pole      \#  9  (   2.0290173521374,   -1.2045224398712)      \\

   amplitude \#  10  (   5.4490417778712D-07,    3.4864743102636D-14)      \\
   pole      \#  10  (   2.6347983285165,  0.)      \\

       Rational approximation degree=   10    \\

       Damping parameter=   -5.00, Expansion point $p_0$:    1.20 \\
          \\

Stable poles:     \\
 
   amplitude \#  1 =   (   200.44842041705,   -6.4297285954932D-08)      \\
   pole      \#  1 =   (  -51.104140221450,  0.)      \\
 
   amplitude \#  2 =   (  1000.06460142389,   -5.7036189955559D-06)      \\
   pole      \#  2 =   (  -5.3490716102094,  0.)      \\

   amplitude \#  3 =   (   499.40738201272,    1.3497166341200D-05)      \\
   pole      \#  3 =   ( -0.50758978101004,  0.)      \\

   amplitude \#  4 =   (   798.94082988133,    2.3349965623979D-05)      \\
   pole      \#  4 =   ( -0.26205674432617,  0.)      \\

   amplitude \#  5 =   (   301.65857589825,    9.1799870303999D-06)      \\
   pole      \#  5 =   ( -0.15111811290477,  0.)      \\

\subsection{Oscillating and decaying exponentials}
5      \\
(400.000,0.)      \\
(175.000,0.)      \\
(175.000,0.)      \\
(-500.000,0.)      \\
(250.000,0.)      \\
(  -2.0000000000000D-03,  0.)      \\
(  -8.0000000000000D-03,    3.0000000000000D-02)      \\
(  -8.0000000000000D-03,   -3.0000000000000D-02)      \\
(  -2.0000000000000D-02,  0.)      \\
( -0.20000000000000,  0.)      \\

    Stability radius      =     10$^{-2}$     \\
    Number of stable poles =  5     \\

     Stable poles and amplitudes:      \\
   amplitude \#  1 =   (   250.00011450201,    9.2737811992658D-07)      \\
   pole      \#  1 =   ( -0.20001227827931,  0.)      \\

   amplitude \#  2 =   (  -499.99976160965,   -8.0435000145924D-04)      \\
   pole      \#  2 =   (  -2.0047256715262D-02,    4.3624394471021D-13)      \\

   amplitude \#  3 =   (   175.00010854502,   -8.5323335360501D-05)      \\
   pole      \#  3 =   (  -8.0240426328942D-03,   -2.9980187019490D-02)      \\

   amplitude \#  4 =   (   175.00006247227,    2.3514996070493D-04)      \\
   pole      \#  4 =   (  -8.0240426329283D-03,    2.9980187019495D-02)      \\

   amplitude \#  5 =   (   399.99945705790,    6.6704669352807D-04)      \\
   pole      \#  5 =   (  -1.9888744154064D-03,  0.)      \\

\subsection{Experimental example borrowed from fast spectroscopy}
The example below Fig.1 shows an experimental noisy signal borrowed from
fast laser spectroscopy. After preprocessing the data and estimating 
the value of the noise present in the data, we perform the Pad\'e-Laplace analysis and
 the results are displayed in Fig.2.
The difficulty in fitting this example is due not only to the presence of noise
but also of the large initial value in the data. Despite these difficulties,
G-MAPLE was able to extract the exponential arguments and corresponding amplitudes
successfully for a rough value of $p_0$. Slightly changing $p_0$ gave a minor
improvement of the fitting. The obstacle in finding a better fit stems from the
fact  it is harder to find a good expansion point $p_0$ for real noisy data.  
The challenge is to find a better  automatic procedure for the determination
of $p_0$ in the case of noisy data. In that case, least-squares \cite{cadzow}
or fast least-squares estimation methods provide better answers \cite{demeure}. 

{\bf Acknowledgements} \\
The  author wishes to acknowledge helpful discussions with 
Randy Borle and Art Monk that helped improve substantially the graphical user
interface of the software. He also thanks Daniel Houde for kindly providing 
two fast spectroscopy data sets for testing the software, and P. Rasolofosaon
for providing reprints/preprints of his work.\\

\begin{figure}[htbp]
\begin{center}
\scalebox{0.7}{\includegraphics[angle=0]{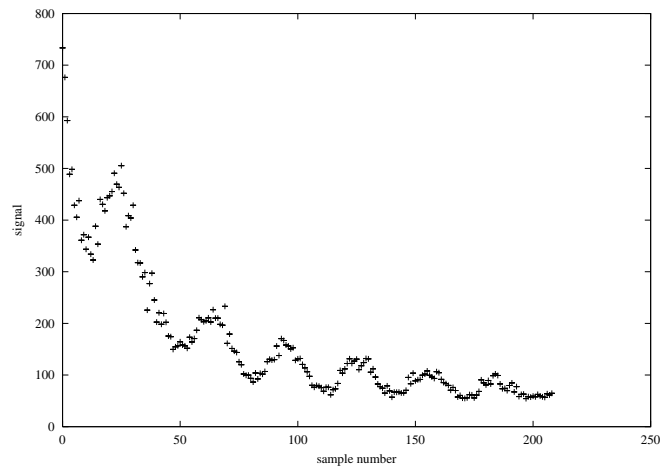}}
\caption{Fast spectroscopy raw data set \# 67}
\end{center}
\label{fig1}
\end{figure}

\begin{figure}[htbp]
\begin{center}
\scalebox{0.7}{\includegraphics[angle=0]{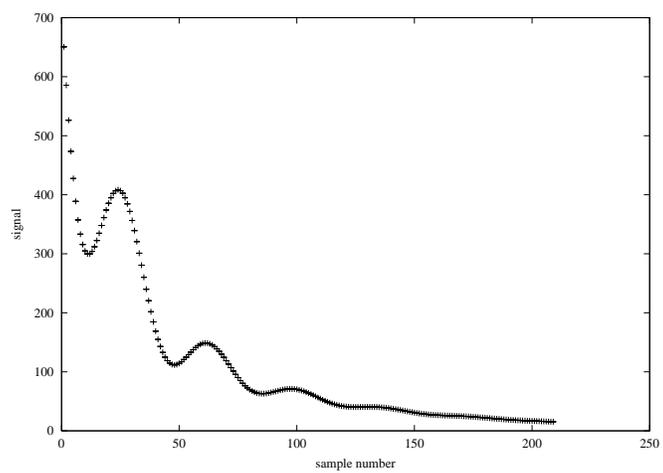}}
\caption{G-Maple result for the fast spectroscopy data set \# 67}
\end{center}
\label{fig2}
\end{figure}

\end{document}